\documentclass[12pt]{article}
\usepackage{sc3conf}
\usepackage{amsfonts}
\usepackage{epsfig}

\newcommand{\text}[1]{\quad\mbox{#1}\quad}
\newcommand{\spr}[2]{\vec{#1}\cdot\vec{#2}}

\newcommand{\pd}[1]{\partial_{#1}}

\begin{document}
\raggedbottom

\title{On the nature of the Blandford-Znajek mechanism}

\authors{Komissarov S.S.}

\addresses{Department of Applied Mathematics, University of 
Leeds, Leeds LS2 9JT, UK.\\ E-mail: serguei@maths.leeds.ac.uk} 
\maketitle

\begin{abstract} 
It is widely accepted in the astrophysical community that 
the event horizon plays crucial role in the 
Blandford-Znajek mechanism of extraction of rotational 
energy of black holes. In fact, this view is a quintessence of the 
Membrane Paradigm of black holes which suggests that the event horizon, 
or rather the so-called stretched horizon, is similar to a rotating 
conducting sphere of finite resistivity. In this paper we argue that
this interpretation is rather misleading and the proper explanation 
of the Blandford-Znajek mechanism has to be sought in    
the properties of the ergospheric region of black holes.      
\end{abstract}

\section{Blandford-Znajek solution and the Membrane Paradigm}

We start with describing some basic properties of
steady-state force-free magnetospheres of black-holes as 
discovered in [1] using the 3+1 formulation of electrodynamics. 
First we introduce vectors of electric and magnetic field
as follows 
\begin{equation}
  E_i = F_{it}  \qquad B^i=\frac{1}{2} 
  e^{ijk} F_{jk}, 
\label{b1}
\end{equation}
where $F_{\mu\nu}$ is the electromagnetic field tensor,and $e$ 
is the Levi-Civita alternating tensor. 
For steady-state axisymmetric solutions  
$E_\phi = 0$. Consider a polar flux tube of magnetic flux  
$$
    \Phi=\int B^i dS_i   
     \qquad ( dS_i=e_{ijk} dx^j_{(1)} dx^k_{(2)} )
$$ 
The angular velocity of magnetic field lines defined as 
\begin{equation}
    \Omega = E_r/\sqrt{\gamma}B^\theta = 
    -E_\theta/\sqrt{\gamma}B^r.  
\label{b2}
\end{equation}
is constant along the field lines and, thus, is a function 
of $\Phi$. Another constant, 
$B_T=\sqrt{-g}F^{r\theta}$, is loosely called the ``azimuthal field''   
because in the Boyer-Lindquist coordinates 
\begin{equation}
    B_T = \alpha g_{\phi\phi} B^\phi
\label{b3}
\end{equation}
where $\alpha$ is the lapse function. 
In addition, neither electric charge nor energy or angular momentum
can flow across the wall of the flux tube and, thus, the total poloidal 
electric current, $I$, energy flux, $\cal{E}$, and angular 
momentum flux, $L$, also depend only on $\Phi$. Moreover,     
\begin{equation}
\frac{d\cal{E}}{d\Phi} = -B_T \Omega,  \qquad
\frac{dL}{d\Phi} = -B_T . 
\label{b4}
\end{equation}

In the original paper by Blandford and Znajek [1] the so-called 
``horizon boundary condition''seems to play a rather important role. 
This condition is a derivative of the usual regularity condition --    
a free-falling observer crossing the horizon should register a finite 
electromagnetic field. This leads to    

\begin{equation}
   B_T = -f(\theta)(\Omega_h-\Omega) \pd{\theta} \Phi
\label{b5}
\end{equation}
at the horizon [2], where $\Omega_h=a/(r_+^2+a^2)$ is the angular 
velocity of the black hole and $f(\theta)$ is a positive function of 
the polar angle. The most important result of [1] is  
the perturbative solution (we shell refer to it as the BZ solution) 
for a monopole-like magnetosphere of a 
slowly rotating black hole that satisfies the ``horizon boundary 
condition'' and  matches Michel's flat space wind solution [3] at 
infinity. 
In such magnetosphere, all magnetic field lines originate 
from the event horizon and rotate with $ \Omega = 0.5 \Omega_h$ 
ensuring outgoing fluxes of energy and angular momentum. 

Although the physical conditions utilized in this model 
are rather obvious, they are the Kerr metric, negligibly small 
mass density of matter, and high conductivity of the magnetospheric 
plasma, no clear explanations of the internal ``mechanics'' of 
the BZ process seems to have been given so far. 
Instead, a surrogate ``explanation'' was 
put foward where the so-called ``stretched 
horizon'' located just outside of the real event horizon,  
is considered as an analogue 
of a magnetized non-perfectly conducting sphere rotating 
in flat space-time with $\Omega_h$ and, thus, inducing  
rotation of magnetic field lines originated from its surface. 
This analogy is often used to describe how the BZ 
mechanism operates to wider audience not particularly familiar 
with general relativity.  It has been pushed all the way in the 
Membrane Paradigm [4] where it is regarded as capturing 
all important features of the real object and even allowing 
rather accurate analysis without resorting to difficult 
general relativistic calculations. In fact, this analogy is 
so popular that in the minds of many the Membrane Paradigm 
and the BZ mechanism are inseparable.  However, nobody can deny 
that without full understanding of the real thing no analogy 
can be relied upon with confidence. As far as the electrodynamics of 
black is concerned, this issue has been addressed in   
in the critical analysis of the BZ mechanism by Punsly and Coroniti [5] 
(see also [6]), who showed that the stretched horizon can not 
communicate with the outgoing electromagnetic wind by means 
of Alfv\'en waves in contrast to a rotating conducting 
sphere in flat space-time.  In [5,6] this causality argument 
is used to revise the role of the event horizon  and even 
to rise questions about stability  of the BZ 
solution.  Recent numerical simulations [13] convincingly show, 
however, that the BZ solution is asymptotically stable and, 
thus, leave only one suspect -- the Membrane Paradigm.

\section{The origin of rotation}

In this section we will try to identify what drives the 
rotation of magnetospheres of black holes. 
For the sake of argument we assume that 
magnetic field lines threading the event horizon also thread  
a massive nonrotating shell located at some distance from 
the hole and for this reason do not rotate, $\Omega =0$. 
Without any loss of generality we will consider 
the case of $\Omega_h,\pd{\theta}\Phi>0$. 
Then eq.\ref{b5} requires $B_T<0$ and the second equation 
in (\ref{b4}) implies $L>0$. Thus, vanishing $\Omega$ results in 
torque being applied to the shell due to building up of azimuthal 
magnetic field! The shell can remain nonrotating only if 
there is another torque which cancels 
the torque applied by the black hole. Otherwise both the shell 
and the magnetic field lines will be forced to rotate in the same 
sense as the black hole, $\Omega >0$, resulting in extraction 
of energy from the hole, $\cal{E}$ $>0$ (see the first equation in
(\ref{b4}).) 

This argument applies only to magnetic field lines penetrating 
the event horizon and makes use of the debatable horizon boundary 
condition. What can we tell about the field lines which do not 
penetrate the horizon? Membrane Paradigm gives no reason for 
such field lines to rotate.  
Once more let us to consider a static magnetosphere where $\Omega=0$ 
and, thus, the electric field as defined in (\ref{b1}) vanishes:     
$$
E_i=0, \quad i=\phi,r,\theta.  
$$
However, the local ``zero angular velocity observer'' (ZAMO), which 
is at rest in the space of the BL space-time foliation, 
registers not only magnetic field  
\vskip 0.5cm 
\begin{displaymath}
B^{\hat i}=\sqrt{g_{ii}} B^i
\end{displaymath} 
but electric field as well 
\begin{displaymath}
E^{\hat \phi}=0,
\end{displaymath}
\vskip 0.3cm 
\begin{displaymath}
E^{\hat r}=-\sqrt{\frac{-g_{\phi\phi}}{g_{rr} \cal L}} 
            \Omega_F B^\theta \not=0 ,
\end{displaymath}
\vskip 0.3cm 
\begin{displaymath}
E^{\hat \theta}=\sqrt{\frac{-g_{\phi\phi}}{g_{\theta\theta} \cal L}} 
            \Omega_F B^r \not=0 ,
\end{displaymath}
where 
$$
{\cal L}=g_{tt}g_{\phi\phi}-g_{t\phi}g_{t\phi}. 
$$
This gives 

\begin{equation}
{\hat B}^2 - {\hat E}^2 = g_{\phi\phi}(B^\phi)^2 - 
     \frac{g_{tt}}{\sin^2\theta \Delta}  
     \left( g_{\theta\theta} (B^\theta)^2 +
       g_{rr} (B^r)^2 \right). 
\end{equation}
High conductivity of black hole magnetospheres (e.g.[1,7]) ensures  
\begin{equation}
   {\hat B}^2-{\hat E}^2 \ge 0. 
\end{equation}

Since $g_{ii}>0$ for all $i$, $\Delta = r^2-2r+a^2$ is positive outside 
of the horizon, and  
$$
    \left\{ \begin{array}{ll} 
          g_{tt} <0 & \text{outside of the ergosphere} \\
          g_{tt} >0 & \text{inside of the ergosphere}
          \end{array} \right.
$$
there must be a nonvanishing azimuthal component of magnetic field,  
$B^\phi$, for all magnetic field lines penetrating the ergosphere
irrespective of whether they penetrate the horizon or not! 
As we have already discussed this leads to torque and ultimately 
ensures rotation of these field lines.  

The key property of the ergospheric region of black holes is 
the inertial frame dragging effect. Relative to any physical 
observer the spacial grid of the Boyer-Lindquist coordinates 
is moving superluminally inside the the ergosphere and subluminally 
outside. Thus, inside the ergosphere the magnetic field with 
$\Omega=0$ is static relative to a superluminally moving spacial 
grid! Moreover, relative to ZAMO the BL spacial grid is rotating in 
the azimuthal direction and it is the azimuthal component of 
magnetic field as measured in the BL coordinate basis which is 
generated due to high conductivity of plasma. This fact suggests 
that the superluminal motion of the spacial grid is very important. 
The following flat space-time example confirms this conclusion. 

Consider an inertial frame in Minkowskian space-time with  
pseudo-Cartesian coordinates $x^{\hat \mu}$. The corresponding 
metric form is 
$$
   ds^2=-(dx^{\hat 0})^2+(dx^{\hat 1})^2+(dx^{\hat 2})^2+
   (dx^{\hat 3})^2
$$
Now let us introduce a rectangular grid moving in the $x^1$-direction 
with velocity $\beta$: 
 
\begin{displaymath}
x^\mu = A^\mu_{\hat\nu} x^{\hat\nu } \text{where} 
 A^\mu_{\hat\nu}=\left(\begin{array}{cccc} 
1 & 0 & 0 & 0 \\  
-\beta & 1 & 0 & 0 \\  
0 & 0 & 1 & 0 \\  
0 & 0 & 0 & 1   
\end{array}\right) .
\end{displaymath}
Notice that this is not a Lorentz transformation. In these new 
coordinates the metric form 

$$
   ds^2=(-1+\beta^2)(dx^0)^2+2\beta dx^0 dx^1 
         +(dx^1)^2+(dx^2)^2+(dx^3)^2 
$$
has a number of properties similar to the Kerr metric in the BL
coordinates. In particular, $x^0$ becomes space-like for 
$\beta^2>1$.  Consider magnetic field static relative 
to the moving grid, which implies $E_i=F_{i0}=0$.   
Then, the electromagnetic field registered by our inertial 
observer is  

$$
B^{\hat i} = B^i, \quad
E_{\hat 1} = 0, \quad 
E_{\hat 2} = \beta B^3, \quad 
E_{\hat 3} = -\beta B^2
$$
which gives  

$$ 
\hat B^2-\hat E^2 = (B^1)^2 +(1-\beta^2)((B^2)^2+(B^3)^2).  
$$

Now we can see that in the case of superluminal motion in 
the $x^1$-direction, $\beta^2>1$, the high conductivity condition, 
$ \hat B^2 - \hat E^2 > 0$, requires $B^1 \not = 0 $. Thus, we observe 
the same behaviour as in the case of black holes.

\section{Causality consideration}

The ``driving force'' of the BZ mechanism 
must be able to communicate with the outgoing wind 
by means of both fast and Alfv\'en waves [5,6,8]. Therefore, it must be
located between the inner and the outer Alfv\'en surfaces 
of a black hole magnetosphere [9] and, thus, lay well outside 
of the horizon (Similar conclusion was reached in [6,10].)

In the limit of force-free degenerate electrodynamics the Alfv\'en 
critical surfaces coincide with the light surfaces given by 
\begin{equation} 
 f(\Omega,r,\theta)= g_{\phi\phi}\Omega^2 +2g_{t\phi}\Omega + g_{tt} = 0, 
\label{a9} 
\end{equation} 
Moreover, for a black hole with positive angular velocity and positive 
outgoing energy flux one has   
\begin{equation} 
  0 < \Omega < \Omega_F     
\label{a16} 
\end{equation}  
for the inner critical surface and 
\begin{equation} 
  0 < \Omega_F < \Omega       
\label{a17} 
\end{equation}  
for the outer critical surface, where $\Omega_F$ is the angular 
velocity of the local ZAMO.
On the surface of the ergosphere 

\begin{equation} 
 f(\Omega,r,\theta)= g_{\phi\phi}\Omega (\Omega -2\Omega_F).
\label{a18} 
\end{equation} 
From this one can see that in the limit $\Omega \rightarrow 0$ the 
inner light surface coincides with the ergosphere, $f$ being positive 
inside and negative outside. Since

\begin{equation} 
 \frac{\partial f}{\partial \Omega}= 2 g_{\phi\phi}(\Omega -\Omega_F) 
  < 0 \quad\mbox{for } \Omega=0, \, \theta \not=0
\label{a19} 
\end{equation} 
the inner critical surface moves inside the ergosphere as $\Omega$ 
increases and must remain inside for all values satisfying (\ref{a16}) 
(The third factor in (\ref{a18}) vanishes only when the outer critical 
surface moves inside the ergosphere.)   

Thus, for all values of $\Omega$ 
consistent with extraction of energy from a black hole, 
there always exists an 
outer region of the ergosphere which is    
causally connected to the outgoing electromagnetic wind.  There is 
no clash between causality and the BZ solution but only between 
causality and the interpretation of this solution given in the 
Membrane Paradigm. 

\begin{figure}[htb]
  \begin{center}
     \includegraphics[bb = 0 0 450 390,
       height=12cm, keepaspectratio]{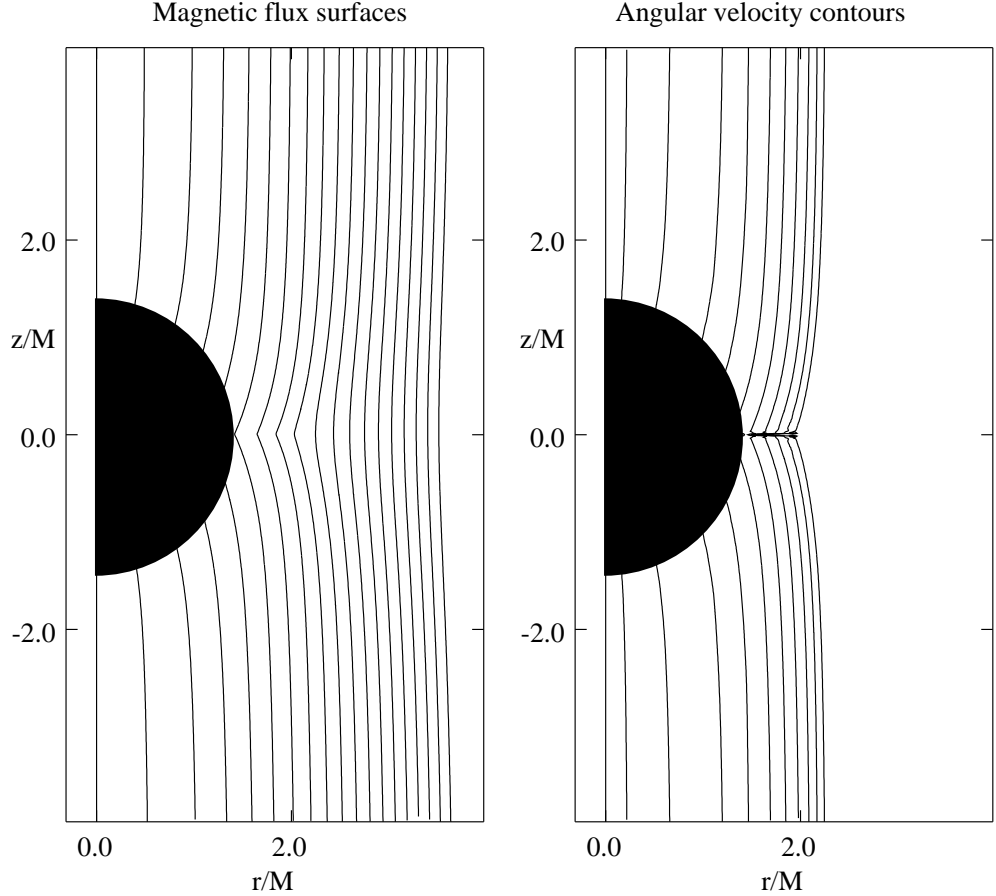}

  \end{center}
\caption{Rotating black hole ($a=0.9M$) in originally uniform magnetic 
field.}
\end{figure}

\section{Numerical simulations}

Consider the problem of a Kerr black hole placed in an originally 
uniform magnetic field aligned along the symmetry axis of the hole. 
In the vacuum solution by Wald [11] there are magnetic field lines 
of three kinds: (i) those that  never enter the ergosphere, (ii) 
those that enter the ergosphere but do not thread the event horizon, 
and (iii) those that thread the event horizon. Should all these  
three types be present in the corresponding solution of force-free
degenerate electrodynamics ([12], ``magnetodynamics'' seems to be 
a better name for this system) we would expect the lines of type (i) 
to remain nonrotating, whereas the lines of both type (ii) and 
type (iii) to rotate. To test this prediction we carried out time-dependent 
numerical 
simulations similar to those described in [13] for the case of monopole 
magnetospheres. We utilized the Kerr-Schild coordinates and placed the 
inner boundary ($r=r_{\mbox{\small in}}$) of the computational domain inside 
the event horizon. The outer boundary ($r=r_{\mbox{\small out}}$) was placed 
far away from the hole to ensure no interference. At $\theta=0$ and 
$\theta=\pi/2$ we used relevant symmetry boundary conditions. 
The initial solution describes the same magnetic field as in [11] and such 
electric field that 1) the degeneracy conditions
 
$$ 
  \spr{E}{B}=0, \quad B^2-E^2>0 
$$          
are satisfied everywhere and 2) the magnetic field lines are non-rotating 
outside of the ergosphere. 

Quite soon after the start of simulations, the second degeneracy 
condition breaks down in the equatorial 
plane inside the ergosphere where magnetic field lines tend to 
rotate slower than the minimum angular velocity of massive particles. 
Even slightly unscreened electric field will create 
strong equatorial current sheet with Compton drag providing required 
resistivity. This electric field will accelerate charged particles to 
rather high Lorentz factors. These energetic particle will pass their 
energy to the background photons which will dump it into the black 
hole. Because for an observer resting at infinity this energy is negative
such interactions result in extraction of rotational energy of 
the black hole. 
However, any significant deviation from $E^2=B^2$ is impossible under the 
typical conditions of astrophysical black holes as it would cause copious 
pair production. In many respects this model of ergospheric dick is 
similar to the one described in [6].    

The proper description of the ergospheric current sheet appears 
to be impossible within the framework of magnetodynamics. However, 
if the net effect of all these processes on the electromagnetic field 
amounts to ensuring the condition $E^2 \approx B^2$ in the equatorial plane 
then the following prescription seems to be quite reasonable. If at the 
end of a normal computational time step we find that in a particular cell
$E^2>B^2$ then the magnitude of $\vec{E}$ is reduced to the magnitude 
of $\vec{B}$. One can show that this corresponds to a source of 
energy and momentum in the cell. 

Figure 1 shows the numerical solution for a black hole with $a=0.9M$ by 
the end of simulations (t=40M) when near the black hole the solution is 
close to a steady state. $M$ is the black hole mass.        
As we expected, all magnetic field lines entering the ergospheric region
are forced to rotate in the same sense as the black hole 
irrespective of whether they thread the horizon or not. 
The magnetic field lines placed outside of the ergosphere remain 
nonrotating with the exception of those passing very close to the 
ergosphere. Their negative angular velocity is most certainly a 
numerical artifact.

\section{Conclusions}

The results presented in this paper strongly suggest 
that the Blandford-Znajek mechanism [1] has  
the same basic ``driving force'' as the Penrose mechanism [14]
which is the ergosphere of a rotating black hole.

\end{document}